\documentclass[12pt]{article}

\usepackage{graphicx}
\usepackage{amssymb}
\usepackage{amsmath}
\usepackage{cite}

\usepackage[default]{jasa_harvard}    
\usepackage{JASA_manu}

\DeclareMathOperator{\sign}{sign}

\DeclareMathOperator{\argmax}{arg\,max}

\begin{document}

\title{Biclustering Via Sparse Clustering}
\author{Qian Liu \\
  Department of Biostatistics \\
  University of North Carolina, Chapel Hill, NC 27599 \\
  email: \texttt{qliu@live.unc.edu} \\
  Guanhua Chen \\
  Department of Biostatistics \\
  University of North Carolina, Chapel Hill, NC 27599 \\
  email: \texttt{guanhuac@live.unc.edu} \\
  Michael R. Kosorok \\
  Department of Biostatistics \\
  University of North Carolina, Chapel Hill, NC 27599 \\
  email: \texttt{kosorok@unc.edu} \\
  Eric Bair \\
  Departments of Endodontics and Biostatistics \\
  University of North Carolina, Chapel Hill, NC 27599 \\
  email: \texttt{ebair@email.unc.edu} \\
}

\maketitle

\newpage

\mbox{}
\vspace*{2in}
\begin{center}
\textbf{Author's Footnote:}
\end{center}
Qian Liu is Doctoral Candidate, Department of Biostatistics,
University of North Carolina, Chapel Hill, NC 27599 (e-mail:
qliu@live.unc.edu). Guanhua Chen is Doctoral Candidate, Department of
Biostatistics, University of North Carolina, Chapel Hill, NC 27599
(e-mail: guanhuac@live.unc.edu). Michael R. Kosorok is W. R. Kenan, Jr. Distinguished Professor and Chair,
Department of Biostatistics, University of North Carolina, Chapel
Hill, NC 27599 (e-mail: kosorok@unc.edu). Eric Bair is Research
Assistant Professor, Department of Biostatistics, University of North
Carolina, Chapel Hill, NC 27599 (e-mail: ebair@email.unc.edu). This
work was partially supported by NIH/NIDCR grants U01DE017018 and
R03DE023592, NIH/NCATS grant UL1RR025747, NIH/NIEHS grant P03ES010126,
and NIH/NCI grant P01 CA142538.

\newpage
\begin{center}
\textbf{Abstract}
\end{center}
In many situations it is desirable to identify clusters that differ
with respect to only a subset of features. Such clusters may represent
homogeneous subgroups of patients with a disease, such as cancer or
chronic pain. We define a bicluster to be a submatrix $\emph{U}$ of a
larger data matrix $\emph{X}$ such that the features and observations
in $\emph{U}$ differ from those not contained in $\emph{U}$. For
example, the observations in $\emph{U}$ could have different means or
variances with respect to the features in $\emph{U}$. We propose a
general framework for biclustering based on the sparse clustering
method of \citeasnoun{witten2010framework}. We develop a method for
identifying features that belong to biclusters. This framework can be
used to identify biclusters that differ with respect to the means of
the features, the variance of the features, or more general
differences. We apply these methods to several simulated and
real-world data sets and compare the results of our method with
several previously published methods. The results of our method
compare favorably with existing methods with respect to both
predictive accuracy and computing time.

\vspace*{.3in}


\newpage

\section{Introduction}
Unsupervised exploratory methods play an important role in the
analysis of high-dimension low sample size (HDLSS) data, such as
microarray gene expression data. Such data sets can be expressed in
the form of a $ n \times p$ matrix $\emph{X}$, where each row
corresponds to one observation each column corresponds to
a feature. Unsupervised learning is a powerful tool for discovering
interpretable structures within HDLSS data without reference to
external information. In particular, clustering methods partition
observations into subgroups based on their overall feature
patterns. In many situations, these underlying subgroups may differ
with respect to only a subset of the features. Such subgroups could be
overlooked if one clusters using all the features.

Biclustering methods may be useful in situations where clusters are
formed by only a subset of the features. Biclustering aims to identify
sub-matrices $\emph{U}$ within the original data matrix $\emph{X}$.
The results may be visualized as two-dimensional signal blocks (after
reordering the rows and columns) containing only a subset of the
observations and features. For example, in a gene expression data set
collected from cancer patients, there may exist a subset of genes
whose expression levels differ among patients with a more aggressive
form of cancer. Identifying such a bicluster may aid in the treatment
of cancer patients.

We define biclusters as sub-matrices $\emph{U}$ of the original data
matrix $\emph{X}$ such that the observations within $\emph{U}$ are
different from the observations not contained in $\emph{U}$ with
respect to the features in $\emph{U}$. In other words, the choice of
features influences which observations form the biclusters. In
general, we can view clustering as a one-dimensional partitioning
method that partitions only the set of observations. Biclustering, on
the other hand, is a two-dimensional partitioning method that
identifies partitions with respect to both features and
observations. However, given a set of features, the problem of
biclustering reduces to the problem of partitioning the observations
with respect to this set of features, a problem which can be solved
using conventional clustering methods. Thus, one may identify
biclusters by identifying the features that define the biclusters and
then clustering with respect to these features. In recent years
several methods have been proposed for identifying features that
define such clusters. We will show how the ``sparse clustering''
method of \citeasnoun{witten2010framework} may be used to identify
biclusters under this framework. The proposed method can be used to
detect biclusters with heterogeneous means and/or variances as well as
more complex differences. We compare our algorithms with some other
existing biclustering approaches by applying the methods to a series of
simulation studies and biomedical data sets.

\section{Methods} \label{S:methods}
\subsection{Sparse Clustering}
The standard $k$-means clustering algorithm partitions a data set into
$k$ sub-categories by maximizing the between cluster sum of squares
(BCSS). The BCSS is calculated by taking the sum of the BCSS's for
each individual feature. This implies that all features are equally
important. However, in many situations the clusters differ with
respect to only a fraction of the features. In such situations, giving
equal weight to all features when clustering may produce inaccurate
results. This is especially true for HDLSS problems. To overcome this
problem, \citeasnoun{witten2010framework} proposed a novel clustering
method which they called ``sparse clustering.'' Under sparse
clustering, each feature is given a nonnegative weight $w_j$, and the
following weighted version of the BCSS is maximized:
\begin{equation} \label{E:sclust_crit}
\begin{split}
\text{maximize}_{C_1, \ldots, C_K, \mathbf{w}} \left\{ \sum_{j=1}^p w_{j} \left( \frac{1}{n} \sum_{i=1}^n \sum_{i'=1}^n d_{i, i', j} - \sum_{k=1}^K \frac{1}{n_k}\sum_{i,i' \in C_k} d_{i,i',j} \right)  \right\} \\
\text{subject to } ||\mathbf{w}||^2 \le 1, ||\mathbf{w}||_1 \le s, w_j \ge 0 \, \forall \, j.
\end{split}
\end{equation}
Here $X_{ij}$ represents observation $i$ for feature $j$ of the data
matrix $\emph{X}$ and $i \in C_k$ if and only if observation $i$
belongs to cluster $k$. $d_{i,i',j}$ is a distance metric between any
pair of observations in $\emph{X}$ with respect to feature $j$. For
$k$-means clustering, we take $d_{i,i',j}=(X_{ij}-X_{i'j})^2$.

\citeasnoun{witten2010framework} describe an iterative procedure for
maximizing (\ref{E:sclust_crit}):
\begin{enumerate}
\item Initially let $w_1=w_2= \ldots w_p$.
\item Maximize (\ref{E:sclust_crit}) with respect to $C_1, C_2,
  \ldots, C_K$ by applying the standard $k$-means algorithm with the
  appropriate weights. In other words, apply the $k$-means algorithm
  where the dissimilarity between observations $i$ and $i'$ is defined
  to be $\sum_{j=1}^p w_j d_{i,i',j}$. \label{en:max_C}
\item Maximize (\ref{E:sclust_crit}) with respect to the $w_j$'s by
  letting
  \begin{equation} \label{E:weight_update}
    w_j = \frac{S(b_j, \Delta)}{\|S(b_j, \Delta)\|_2}
  \end{equation}
  Here $b_j$ is the (unweighted) between cluster sum of squares for
  feature $j$ and $S(x, y) = \sign(x)(|x|-y)_+$ is a soft-threshold
  operator. $\Delta$ is chosen so that $\sum_j |w_j|=s$ ($\Delta=0$ if
  $\sum_j |w_j| \leq s$). See \citeasnoun{witten2010framework} for the
  justification for (\ref{E:weight_update}). \label{en:max_w}
\item Iterate steps \ref{en:max_C} and \ref{en:max_w} until the
  algorithm converges.
\end{enumerate}
Note that (\ref{E:weight_update}) implies that as $s$ decreases, the
number of nonzero $w_j$'s decreases. Thus, for sufficiently small
values of $s$, only a subset of the features contribute to the cluster
assignments, so this method is useful in situations where the clusters
differ with respect to only a subset of the features.

A variant of this procedure can be used to perform sparse hierarchical
clustering. In sparse hierarchical clustering, each feature is once
again given a weight and the cluster hierarchy is constructed using
these weighted features. The value of the weights again depends on a
tuning parameter, and some weights are forced to 0 when the tuning
parameter is sufficiently small. See \citeasnoun{witten2010framework}
for details.

\subsection{Biclustering Via Sparse Clustering} \label{S:mean_biclust}
As described earlier, the objective of biclustering is to identify
submatrices $\emph{U}$ of a data matrix $\emph{X}$ such that the
observations containing in $\emph{U}$ differ from the observations not
contained in $\emph{U}$ with respect to the features contained in
$\emph{U}$. One possible strategy to identify such biclusters is to
apply 2-means sparse clustering. One could define the observations of
$\emph{U}$ to be the observations in the smaller cluster identified by
the procedure and the features in $\emph{U}$ to be the features with
nonzero weights.

\begin{figure}
\centering
\includegraphics[scale=0.32]{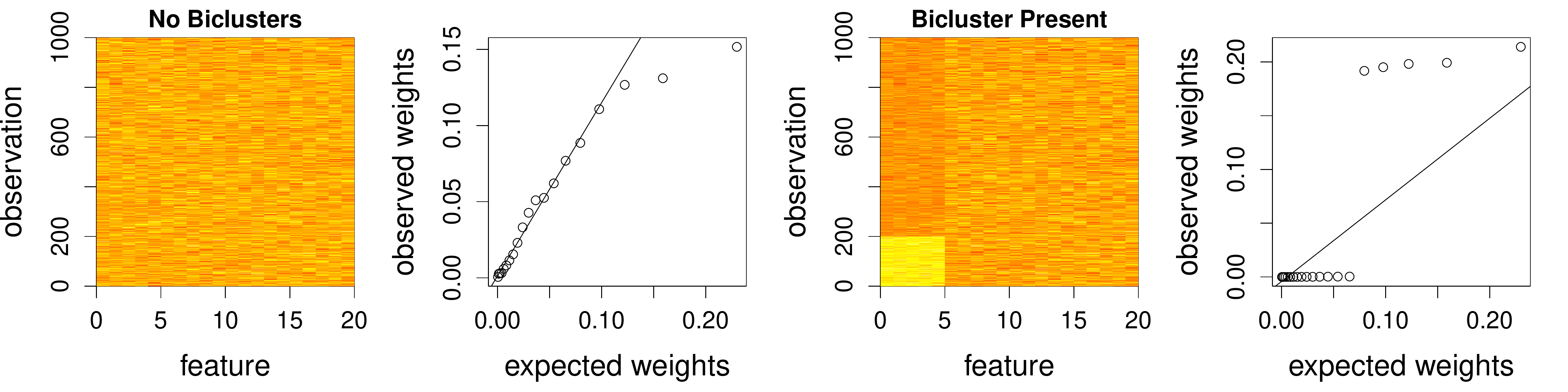}
\caption{\textit{Illustration of the proposed biclustering method.}
  Panels 1 and 3 show heat maps for two (artificial) data sets, and
  panels 2 and 4 show quantile-quantile plots of the the weights
  obtained by sparse clustering on each data set versus the weights
  under the null distribution. The first data set contains no
  biclusters, and the feature weights obtained by our procedure are
  very close to the expected feature weights under the null
  distribution. The second data set contains a bicluster consisting of
  5 features. Note that the weights of these five features are much
  greater than the expected weights under the null distribution and
  the remaining weights will be less than expected under the null
  distribution. See Section \ref{S:null_dist} for an explanation of
  how the expected weights under the null distribution were calculated
  in this illustration.} \label{F:demo_fig}
\end{figure}

The list of features with nonzero weights depends on the tuning
parameter $s$, so this approach to biclustering requires one to choose
the correct value of this tuning parameter. One possible approach for
choosing $s$ is described in \citeasnoun{witten2010framework}, but in
our experience it tends to give nonzero weights to too many
features. Thus, we propose an alternative method for identifying the
features that belong to the bicluster. First, note that if sparse
clustering is applied with $s=\sqrt{p}$, then no soft thresholding
will be performed on the weights and all weights be nonzero. The
motivation for our method is the following: Suppose that sparse
2-means clustering is applied with $s=\sqrt{p}$. Let $w_{(1)},
w_{(2)}, \ldots, w_{(p)}$ denote the order statistics of the weights
produced by the sparse clustering procedure, and let $w_{(1)_0},
w_{(2)_0}, \ldots, w_{(p)_0}$ denoted the expected values of these
order statistics under the null hypothesis that no bicluster
exists. If this null hypothesis is true, then we would expect that
$w_{(j)} \approx w_{(j)_0}$ for all $j$. However, if $m$ features form
a bicluster, then we would expect that $w_{(j)}>w_{(j)_0}$ for $j>p-m$
and $w_{(j)}<w_{(j)_0}$ otherwise. See Figure \ref{F:demo_fig} for an
illustration.

Thus, our proposed biclustering method is described below:
\begin{enumerate}
\item Apply the 2-means sparse clustering algorithm with $s=\sqrt{p}$
  to obtain clusters $C_1$ and $C_2$ and weights $w_1, w_2, \ldots
  w_p$.
\item Perform a Kolmogorov-Smirnov test of the null hypothesis that
  the distribution of $w_1, w_2, \ldots, w_p$ is the same as the
  expected distribution of the weights under the null hypothesis of no
  clusters. \label{en:kstest}
\item If the test in Step \ref{en:kstest} fails to reject the null
  hypothesis, then terminate the procedure and report that no
  biclusters were identified.
\item If the test in Step \ref{en:kstest} rejects the null hypothesis,
  then let
\begin{equation} \label{E:scbiclust_crit}
  m = \argmax_j \left(w_{(p-j+1)}-w_{(p-j+1)_0}\right) -
     \left(w_{(p-j)}-w_{(p-j)_0}\right)
\end{equation}
Intuitively, we are choosing an $m$ such that observation $p-m+1$ is
``above the line'' in Figure \ref{F:demo_fig} and observation $p-m$
is ``below the line.''
\item Return a bicluster containing the $m$ features with the largest
  weights and the observations belonging to either $C_1$ or $C_2$
  (whichever is smaller).
\end{enumerate}
We recommend that the data matrix be normalized such that all features
have mean 0 and standard deviation 1 before applying the procedure. We
will call this procedure ``SC-Biclust'' (an abbreviation for
biclustering based on sparse clustering).

Let $b_j$ denote the between cluster sum of squares for feature
$j$. Suppose that the mean of the observations in $C_1$ is $\mu_{1,j}$
and the mean of the observations in $C_2$ is $\mu_{2,j}$. Then it is
easy to verify that
\begin{equation}
  E(b_j) = 1 + np(1-p)(\mu_{1,j} - \mu_{2,j})^2,
\end{equation}
where $p$ is the probability that a given observation belongs to
$C_1$. This implies that $E(b_j)=1$ if $\mu_{1,j}=\mu_{2,j}$, which
would be the case of feature $j$ does not belong to the
bicluster. However, if $\mu_{1,j} \neq \mu_{2,j}$, then $E(b_j)$ will
increase as $n$ increases. Thus, assuming that $\mu_{1,j} \neq
\mu_{2,j}$ for at least one $j$ (which will always be the case when a
bicluster exists), (\ref{E:weight_update}) implies that $w_j =
b_j/\sqrt{\sum_k b_k^2} \to 0$ as $n$ increases for all $j$ such that
$\mu_{1,j} = \mu_{2,j}$ (i.e., all $j$ that do not belong to the
bicluster). This indicates that the criteria (\ref{E:scbiclust_crit})
is consistent for selecting the features that belong to the bicluster,
assuming that $C_1$ and $C_2$ are correctly identified and the
conditions of the law of large numbers are satisfied.

One may wish to identify secondary biclusters in a data set after
identifying a primary bicluster. One simple approach to identify such
secondary biclusters is described below:
\begin{enumerate}
\item Identify a primary bicluster $U_1$ as described above.
\item Define a matrix $X'$ as follows:
  \begin{equation}
    x_{ij}' =
    \begin{cases}
      x_{ij} &\text{if $x_{ij} \notin U_1$} \\
      x_{ij} - \bar{X}_{U_1,j} + \bar{X}_{U_1',j} &\text{if
        $x_{ij} \in U_1$}
    \end{cases}
  \end{equation}
  Here $\bar{X}_{U_1,j}$ denotes the sample mean of the $j$th feature
  of $U_1$ and $\bar{X}_{U_1',j}$ denotes the sample mean of the
  $j$th feature of the elements of $X$ that are not in $U_1$.
\item Apply the biclustering algorithm to the matrix $X'$.
\end{enumerate}
The above procedure may be repeated as many times as desired to
identify multiple biclusters in the same data sets (although the
procedure should be terminated if it fails to reject the null
hypothesis that no biclusters exist in Step
\ref{en:kstest}).

\subsection{Estimating the Null Distribution of the
  Weights} \label{S:null_dist}
This method requires one to know the expected order statistics of the
weights under the null hypothesis that no clusters exist. If this
distribution is unknown, it may be approximated as follows:
\begin{enumerate}
\item Apply the 2-means sparse clustering algorithm with $s=\sqrt{p}$
  to obtain clusters $C_1$ and $C_2$, as before.
\item Fix $C_1$ and $C_2$ and permute the rows of $X$ to calculate
  weights $w_1^*, w_2^*, \ldots, w_p^*$. \label{en:permute}
\item Repeat Step \ref{en:permute} $B$ times.
\item Approximate $w_{(j)_0}$ as $w_{(j)_0} = \sum_k w_{(j)k}^*/B$,
  where $w_{(j)k}^*$ represents the $j$th order statistic of the
  weights from the $k$th iteration of Step \ref{en:permute}.
\end{enumerate}

This procedure will provide an estimate of the expected values of the
order statistics of the weights, but it is very expensive
computationally for large data sets. It would be desirable to develop
a faster alternative. Fortunately, if the sparse clustering procedure
is modified slightly, the exact distribution of the weights can be
calculated under mild assumptions.

First, note that the criterion in (\ref{E:sclust_crit}) can be written
as $\sum_j w_j b_j$, where $b_j$ is the between cluster sum of squares for
feature $j$. If we modify the procedure to minimize $\sum_j w_j
\sqrt{b_j}$ rather than $\sum_j w_j b_j$, then (\ref{E:weight_update})
implies that the optimal $w_j$'s are given by
\begin{equation} \label{E:sqrt_weights}
  w_j = \frac{\sqrt{b_j}}{\sqrt{\sum_k b_k}}
\end{equation}
assuming $s=\sqrt{p}$ (implying that $\Delta=0$ in
(\ref{E:weight_update})). Now under the null hypothesis that no
clusters exist, there is no difference in the means of the
observations in $C_1$ and $C_2$ for all features, implying that $b_j
\sim \chi^2_1$ for all $j$. Thus, (\ref{E:sqrt_weights}) implies that
$w_j^2$ has a $\text{Beta}(1/2, (p-1)/2)$ distribution if the $b_j$'s
are independent. Thus, if we use this criterion to select the
clusters, we can test the null hypothesis that no bicluster exists by
performing a Kolmogorov-Smirnov test of the null hypothesis that the
$w_j^2$'s have a $\text{Beta}(1/2, (p-1)/2)$ distribution. Similarly,
in (\ref{E:scbiclust_crit}), $w_{(j)_0} = E(\sqrt{B_{(j)}})$, where $B
\sim \text{Beta}(1/2, (p-1)/2)$. Although there is no simple closed
form expression for $E(\sqrt{B_{(j)}})$, it can be easily approximated
numerically. We will use this method to approximate the null
distribution of the weights in all subsequent examples unless
otherwise noted.

\subsection{Variance Biclustering and Other
  Variations} \label{S:var_biclust}
Note that sparse 2-means clustering is only used in the initial step
of our biclustering procedure. In principle any clustering procedure
that produces two clusters could be used in place of sparse 2-means
clustering. Sparse 2-means clustering is an obvious choice to identify
putative biclusters since it is designed to identify clusters that
differ with respect to only a subset of the features. However, in some
situations it may be desirable to use a different clustering procedure
to identify the putative biclusters.

One important application where it may be useful to use an alternative
clustering procedure is variance biclustering. The biclustering method
described in Section \ref{S:mean_biclust} is designed to identify
biclusters whose mean differs from the mean of the observations that
do not belong to the bicluster. In some situations, however, one may
wish to identify biclusters that have unusually high (or low) variance
compared to observations that are not in the bicluster. For example,
when analyzing DNA methylation data, biclusters that exhibit high
variance may reveal possible functional regions in the genome.

To identify variance biclusters, we propose the following simple
modification of 2-means clustering in order to identify clusters whose
variances differ from one another:
\begin{enumerate}
\item Initially assign each observation to either cluster 1 or cluster
  2. \label{en:var_init}
\item For $i=1,2,\ldots,n$, move observation $i$ from cluster 1 to
  cluster 2 (or from cluster 2 to cluster 1) if
  \begin{equation} \label{E:var_crit}
    \sum_{j=1}^p \log(|s_{j,C_1}^2 - s_{j,C_2}^2|+1)
  \end{equation}
  is increased after moving the observation to the other cluster. Here
  $s_{j,C_k}$ represents the standard deviation of feature $j$ for the
  observations in cluster $k$. \label{en:move_obs}
\item Repeat Step \ref{en:move_obs} until the procedure converges.
\end{enumerate}
Note that we did not specify how the initial cluster assignments in
step \ref{en:var_init} were performed. The simplest approach is to
simply assign each observation to a cluster randomly. An alternative
approach is to calculate the variance of the data for each observation
across the features. The observations are then partitioned based on
their variances: half of the observations with the largest variances
are initially assigned to cluster 1 and the other half of the
observations (with the smallest variances) are initially assigned to
cluster 2. Our preliminary work suggests that both approaches produce
comparable results but the latter approach tends to be faster, so we
will use this approach in all subsequent examples.

Also, note that this procedure can be easily modified to consider
feature weights by replacing (\ref{E:var_crit}) with
\begin{equation} \label{E:wvar_crit}
  \sum_{j=1}^p w_j \log(|s_{j,C_1}^2 - s_{j,C_2}^2|+1)
\end{equation}
A sparse version of this algorithm (motivated by the sparse clustering
algorithm) is also possible, as described below:
\begin{enumerate}
\item Initially let $w_1=w_2= \ldots w_p$.
\item Maximize (\ref{E:wvar_crit}) with respect to $C_1$ and $C_2$
  by applying the above procedure with the appropriate
  weights. \label{en:vmax_C}
\item Maximize (\ref{E:wvar_crit}) with respect to the $w_j$'s by
  letting
  \begin{equation}
    w_j = \frac{S(b_j, \Delta)}{\|S(b_j, \Delta)\|_2}
  \end{equation}
  where $b_j = \log(|s_{j,C_1}^2 - s_{j,C_2}^2|+1)$.  \label{en:vmax_w}
\item Iterate steps \ref{en:vmax_C} and \ref{en:vmax_w} until the
  algorithm converges.
\end{enumerate}
By replacing 2-means sparse clustering with the procedure described
above, the biclustering algorithm described in Section
\ref{S:mean_biclust} can be used to identify variance biclusters. If
one wishes to identify secondary variance biclusters, one may define a
matrix $X'$ as follows:
\begin{equation}
  x_{ij}' =
  \begin{cases}
    x_{ij} &\text{if $x_{ij} \notin U_1$} \\
    \frac{x_{ij} \sigma_{U_1',j}}{\sigma_{U_1,j}} &\text{if
      $x_{ij} \in U_1$}
  \end{cases}
\end{equation}
where $\sigma_{U_1,j}$ denotes the standard deviation of the $j$th
feature of $U_1$ and $\sigma_{U_1',j}$ denotes the standard
deviation of the $j$th feature of the elements of $X$ that are not in
$U_1$.

Note that this procedure requires an estimate of the null distribution
of the $w_j$'s. This null distribution may be estimated by permuting
the rows of $X$ as described in Section \ref{S:null_dist}.
Alternatively, one can take advantage of the fact that $n_1
s_{j,C_1}^2 \sim \chi^2_{n_1}$ and $n_2 s_{j,C_2}^2 \sim \chi^2_{n_2}$
for all $j$ under the null hypothesis of no variance biclusters, where
$n_1$ and $n_2$ are the number of observations in $C_1$ and $C_2$,
respectively. The null distribution of the $b_j$'s (and hence the
$w_j$'s) can be estimated by simulating chi-square random variables
and calculating the $w_j$'s for each set of simulated values. We will
use this method to approximate the null distribution in all examples
in this manuscript, since the permutation-based approach is much
slower.

Other variations of this biclustering procedure are possible. For
example, rather than using sparse 2-means clustering to identify the
putative biclusters in the first step of the procedure, one could use
some form of hierarchical clustering and then partition the cluster
hierarchy into two clusters. We will provide a simulated example below
where applying hierarchical clustering with single linkage to identify
the biclusters produces better results than sparse 2-means
clustering.

\subsection{Existing Biclustering Methods} \label{method_review}
A variety of biclustering methods have been proposed. One simple and
commonly used approach is to independently apply hierarchical
clustering to both the rows and columns of a data set
\cite{eisen1998cluster}. Several improvements of this simple approach
have been proposed \cite{getz2000coupled,weigelt2005molecular}. Other
biclustering methods directly search for submatrices $\emph{U}$ such
that the mean of the observations in $\emph{U}$ is higher than the
mean of the observations not in $\emph{U}$. The ``Plaid'' method of
\citeasnoun{lazzeroni2002plaid} approximates a data matrix $X$ as a
sum of submatrices whose entries follow two-way ANOVA models. At each
step of the procedure, the algorithm searches for a submatrix that
maximizes the reduction in the overall sum of squares. Similarly, the
``Large Average Submatrix'' (LAS) method of
\citeasnoun{shabalin2009finding} assumes that the data matrix can be
expressed as a sum of constant submatrices plus Gaussian noise. These
submatrices are identified using an iterative search procedure. Also,
the ``sparse biclustering'' method of  \citeasnoun{tan2013sparse}
assumes that the $n$ observations belong to $K$ unknown and
non-overlapping classes, and the $p$ features belong to $R$ unknown
and non-overlapping classes. The mean value of all the features in
each class is assumed to be the same. Class labels are obtained by
maximizing the log likelihood, and sparsity is obtained by imposing an
\textit{$\ell_1$} penalty on the log likelihood.

Other methods for identifying biclusters utilize the singular value
decomposition (SVD) of the data matrix.
\citeasnoun{lee2010biclustering} propose a method that searches for a
low-rank ``checkerboard-structured'' approximation for a data matrix
by calculating a weighted form of the SVD. An adaptive lasso penalty
\cite{zou2006adaptive} is applied to the weights, forcing both the
left and right singular vectors to be sparse. The nonzero entries in
the resulting (sparse) singular vectors correspond to the observations
and features forming the bicluster. \citeasnoun{chen2013biclustering}
develop a generalization of this method called ``Heterogeneous Sparse
Singular Value Decomposition'' (HSSVD). HSSVD approximates the data as
the sum of a ``mean layer'' and a ``variance layer'' (plus random
noise) and identifies biclusters in these two layers. The inclusion of
a ``variance layer'' allows one to identify variance biclusters as
well as mean biclusters.

While these methods have been useful for many problems, they have
certain shortcomings. As we will demonstrate below, they may fail to
identify biclusters in simple simulations. Also, with the exception of
the HSSVD method, these existing methods can only identify biclusters
whose means differ from the observations not in the bicluster. (HSSVD
can also identify biclusters whose variances differ.) However,
biclustering methods based on the SVD have other shortcomings. These
methods can identify the presence of biclusters but cannot determine
which observations and features belong to the bicluster without using
arbitrary cutoffs.

\subsection{Evaluating the Reproducibility of
  Biclusters} \label{method_cv}
We propose an intuitive method to evaluate the reproducibility of the
biclusters identified by each method. We randomly partition the
original data matrix $X$ into two submatrices $X_1$ and $X_2$, each of
which contains half of the observations. Denote the primary bicluster
identified within $X$ as $U$, and let $U_1$ and $U_2$ be the primary
biclusters within $X_1$ and $X_2$, respectively. We treat $U$ as the
reference or the ``correct'' bicluster, and record four rates:
\textbf{1)} The percentage of observations that are
misclassified (i.e. the percentage of observations that are either in
$U_1$/$U_2$ but not $U$ or in $U$ but not in $U_1$/$U_2$); \textbf{2)}
The percentage of false negatives (i.e. the average percentage of
features in $U$ that are not in $U_1$/$U_2$); \textbf{3)} The
percentage of false positives (i.e. the average number of features in
$U_1$/$U_2$ that are not in $U$); and \textbf{4)} The percentage
of features that are misclassified (i.e. features that are identified
as significant on sub-matrix $U_1$ but not $U_2$, or vice
versa). We repeated the procedure 10 times on each simulated data set
and averaged over the 10 iterations.

\subsection{Computational Details}
In the later sections, we will compare our proposed biclustering
algorithm with several existing methods, specifically Plaid, LAS,
SSVD, and HSSVD. The Plaid algorithm was implemented in the R package
``biclust.'' The default setting was used, and the data sets
were feature/column scaled before running through the algorithm. The
LAS algorithm is available at https://genome.unc.edu/las/. The
default settings were used, including the data transformation step if
recommended by the method. The SSVD functions are available at
http://www.unc.edu/$\sim$haipeng/, and the HSSVD functions can be
found at http://impact.unc.edu/impact7/HSSVD. Again, the
default settings were used for both methods. Note that this
implementation of HSSVD does not use a pre-specified rank. For easy
visualization of these two SVD based methods, we transformed the
resulting singular vectors/matrices to be 0/$\pm1$ based on the signs
of the entries making the plots. When comparing the prediction
accuracy and reproducibility of these methods, we further dichotomized
the results as 0/1, since we only care about whether an object or
feature is inside the sub-matrix $U$. The sparse biclustering
algorithm was implemented using the ``sparseBC'' R package with the
default settings. We used $K=2$ and $R=2$ to force the sparse
biclustering algorithm to identify only a single bicluster so that its
results can be compared with other methods that identify one bicluster
at a time. Our proposed method was implemented using a modified
version of the ``sparcl'' R package. All calculations were performed
using a single core of a 2.66 GHz Intel Core 2 Quad processor on a
Linux-based system.

\section{Results}
\subsection{Simulation Studies}
We first evaluated the performance of our method on a variety of
simulated data sets and compared its performance with the
biclustering methods described in Section \ref{method_review}. The
methods were compared with respect to computing time, prediction
accuracy, and reproducibility (defined in Section
\ref{method_cv}). To evaluate the prediction accuracy when identifying
a single bicluster, we compared three rates: observation
misclassification rate, feature false positive rate (FPR), and feature
false negative rate (FNR).

For the sequential biclustering simulations (simulations 3 and 5), the
identification of the current bicluster depends on all the biclusters
identified previously and there is no ``correct'' sequence for
identification. Thus, we recorded the prediction accuracy in a
different manner. Specifically, when there existed two biclusters, the
reasonable result would be the identification of either bicluster 1 or
2, or a larger bicluster that covers both the signal blocks, which
will be referred to as ``bicluster 1+2.'' For each simulation, we
determined which of the three biclusters was identified by each
method. For each method, we recorded the percentage of simulations
when each of the three possible biclusters was identified. Also,
instead of comparing the mismatch rates for observations and features
separately, we recorded the FPR and FNR of the entries. The
reproducibility analysis described in Section \ref{method_cv} was only
performed on simulation studies 1, 2, and 4 for computational
reasons.

Some methods failed to produce ``valid'' biclusters for some of the
simulated data sets. We defined a ``valid'' bicluster to be a
bicluster consisting of at least two observations and two
features. For each simulation, the number of invalid results over the
100 simulations was tabulated, and invalid results were not used when
calculating the average accuracy of each method. Finally, we examined
the number of biclusters identified by each method for simulations 1
and 3 and compared it to the number of biclusters that truly exist.

\subsubsection{Primary Bicluster Identification} \label {prim_biclust}
\begin{figure}
\centering
\includegraphics[scale=0.32]{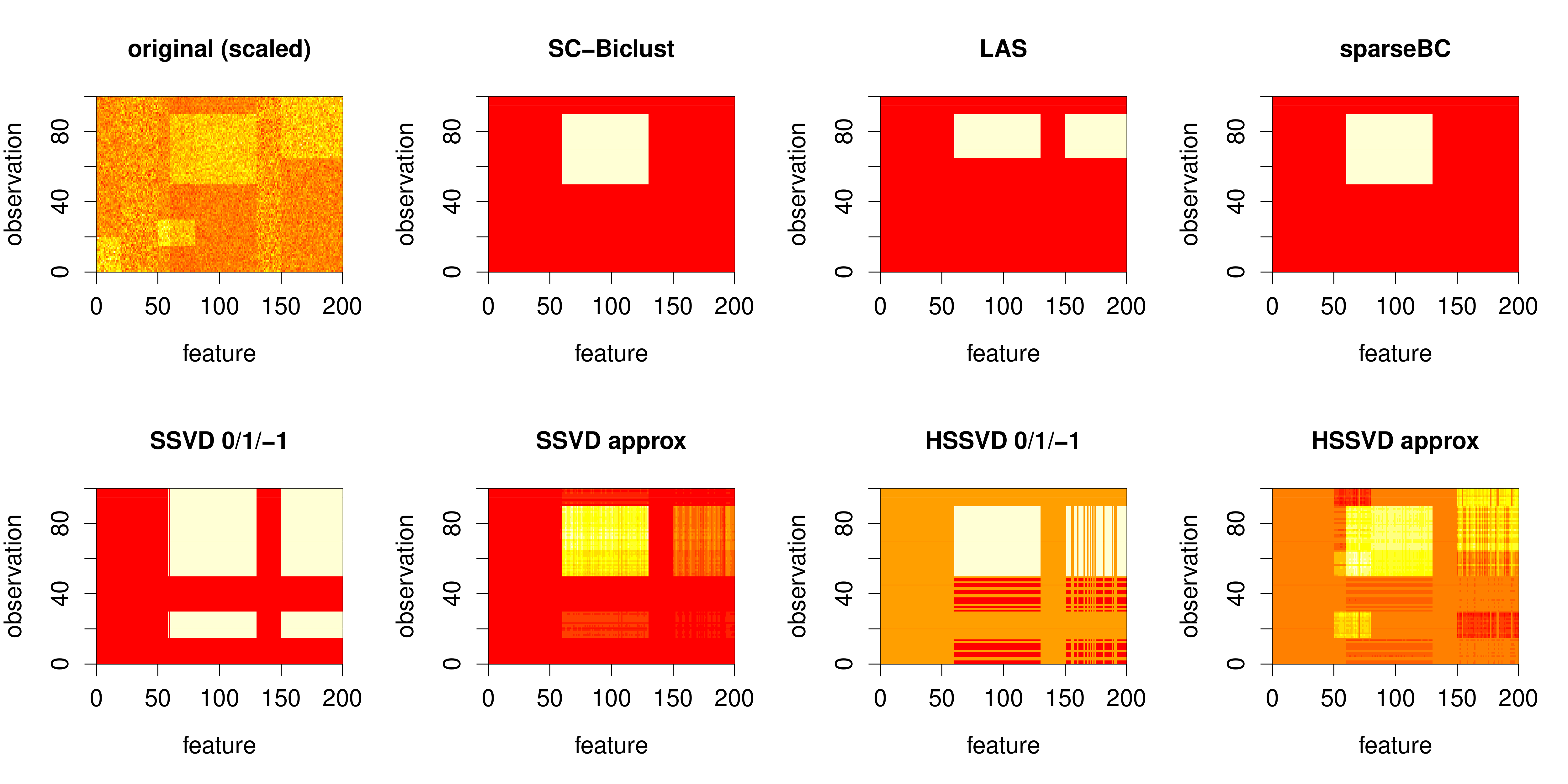}
\caption{\textit{Simulation example: primary bicluster
  identification.} This is an illustration of a single simulation
  from the first simulation scenario. The first panel shows a heat map
  of the (scaled) data. The primary bicluster is the rectangular
  yellow block in the middle. The remaining panels show the biclusters
  identified by SC-Biclust, LAS, sparse biclustering, SSVD, and HSSVD,
  with the white regions corresponding to the biclusters. For SSVD and
  HSSVD, both the 0/1/-1 indicator matrix and the approximation matrix
  are plotted.} \label{F:sim_norm}
\end{figure}

~\\
In this study, each simulated data set contained four non-overlapping
bicluster signals generated from normal distributions, and the
comparison was focused on identifying the primary bicluster. Each
simulated data set comprised a 100 $\times$ 200 matrix with
independent entries where each column represents a feature and each
row represents an observation. The background entries followed a
standard normal distribution with mean 0 and standard deviation 1. We
denote the distribution as $N(0, 1)$, where $N(a, b)$ represents a
normal random variable with mean $a$ and standard deviation $b$. The
four non-overlapping rectangular shaped biclusters were constructed in the
following manner: bicluster 1, consisting of observations 1-20 and
features 1-20 (denoted as [1-20, 1-20]) added a $N(2, 1)$ layer to the
background, bicluster 2 [16-30, 51-80] added a $N(3, 1)$ layer to the
background, bicluster 3 [51-90, 61-130] added a $N(3, 1)$ layer to the
background, and bicluster 4 [66-100, 151-200] added a $N(2, 1)$ layer
to the background. Bicluster 3 was the primary bicluster, since it was
the largest bicluster and had the largest mean difference from the
background, so we expected the algorithms to detect this bicluster as
the first layer. Figure \ref{F:sim_norm} shows the biclustering
results from one of the simulations. Under the given data structure,
the Plaid algorithm failed to identify any biclusters for all the
simulations. Each simulated data set was partitioned as described in
Section \ref{method_cv} to evaluate the reproducibility of the
biclusters.

\subsubsection{Departure from Normality}
\begin{figure}
\centering
\includegraphics[scale=0.35]{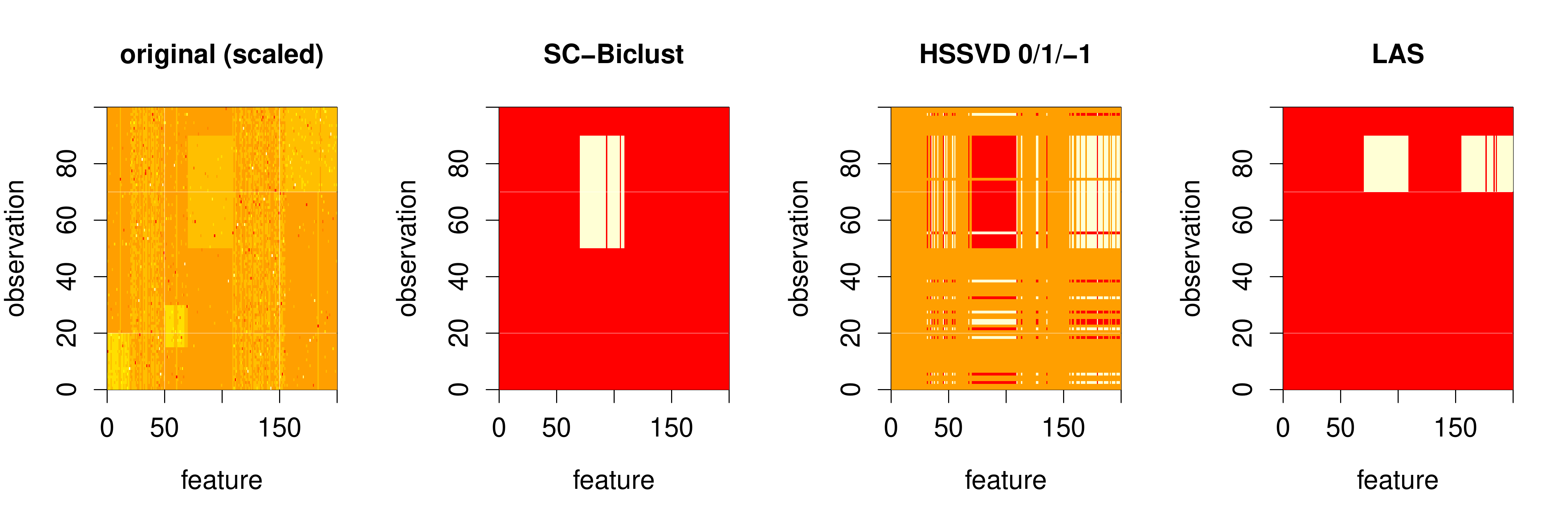}
\caption{\textit{Simulation example: departure from normality.} This
  is an illustration of a single simulation from the second simulation
  scenario. The first panel shows a heat map of the (scaled) data. The
  primary bicluster is the rectangular yellow block in the middle. The
  remaining panels show the biclusters identified by SC-Biclust,
  HSSVD, and LAS, with the white regions corresponding to the
  biclusters.} \label{F:sim_cauchy}
\end{figure}

~\\
In this study, we simulated data sets with four non-overlapping
bicluster signals similar to the data sets that were simulated in
Section \ref{prim_biclust}. The main difference is that the data were
generated from Cauchy distributions with infinite moments. Each
simulated data set comprised a 100 $\times$ 200 matrix with
independent entries. The background entries followed a Cauchy
distribution with location shift $0$ and scale $1$. We denote
the distribution as $\text{Cauchy}(0, 1)$, where $\text{Cauchy}(a, b)$
represents a Cauchy random variable with location shift $a$ and scale
$b$. The four non-overlapping rectangular shaped biclusters were
constructed in the following manner: bicluster 1 [1-20, 1-20] added a
$\text{Cauchy}(75, 1)$ layer to the background, bicluster 2 [16-30,
51-80] added a $\text{Cauchy}(50, 1)$ layer to the background,
bicluster 3 [51-90, 71-110] added a $\text{Cauchy}(200, 1)$ layer to
the background, and bicluster 4 [71-100, 156-200] added a
$\text{Cauchy}(75, 1)$ layer to the background. Bicluster 3 was the
primary bicluster, and we expected the algorithms to detect this
bicluster as the first layer. Figure \ref{F:sim_cauchy} compares the
biclustering results of each method for one of the simulations. Each
simulated data set was partitioned as described in Section
\ref{method_cv} to evaluate the reproducibility of the biclusters.

\subsubsection{Sequential Biclusters with Overlap}
\begin{figure}
\centering
\includegraphics[scale=0.31]{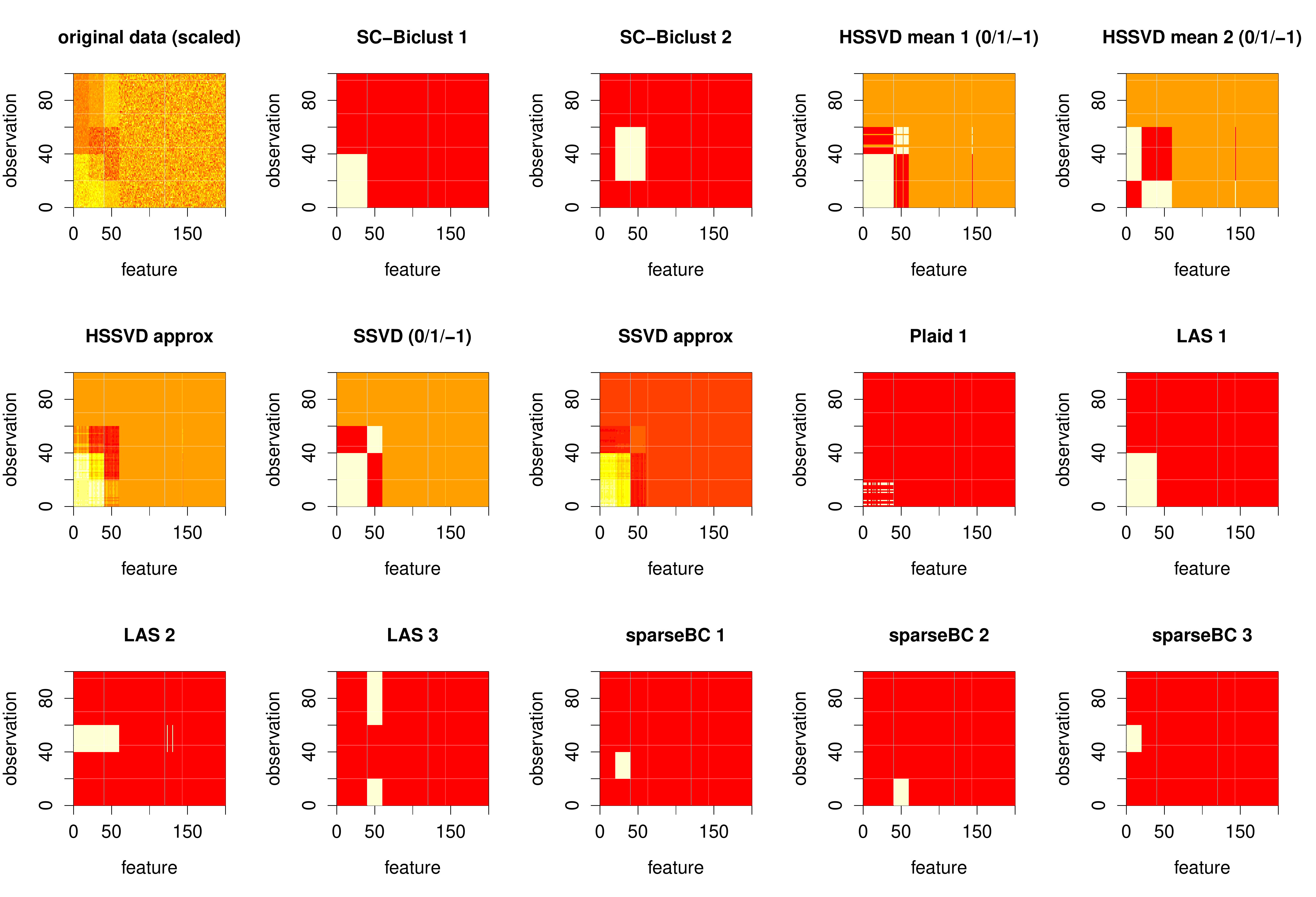}
\caption{\textit{Simulation example: sequential biclusters with
  overlap.} This is an illustration of a single simulation from the
  third simulation scenario. The first panel shows a heat map of the
  (scaled) data. The two overlapping biclusters are in the bottom left
  corner of the data matrix; one is in red and the other is in yellow.
  The remaining panels show the first two biclusters identified by
  SC-Biclust and HSSVD, the first bicluster identified by SSVD and
  Plaid, and the first
  three biclusters identified by LAS and sparse biclustering. The white
  regions correspond to the biclusters. For SSVD and HSSVD, both the
  0/1/-1 indicator matrix layers and the overall approximation
  matrices are plotted.} \label{F:sim_overlap}
\end{figure}

~\\
In this study, we simulated data sets with overlap between two
biclusters. Each simulated data set comprised of two layers, each of
which was a 100 $\times$ 200 matrix with independent entries. The
background data (i.e., observations that do not belong to the
bicluster) were $N(0, 0.5)$. The first layer contained a bicluster
[1-40, 1-40] generated from $N(7, 2)$, and the second layer contained
a bicluster [21-60, 21-60] generated from $N(-5, 3)$. The final data
set was the sum of the two layers. Note that observations 21-40 and
features 21-40 are contained in both biclusters. Figure
\ref{F:sim_overlap} shows the biclustering results from one of the
simulations. Reproducibility of the biclusters was not evaluated for
this simulation scenario.

\subsubsection{Non-Spherical Biclusters}
\begin{figure}
\centering
\includegraphics[scale=0.35]{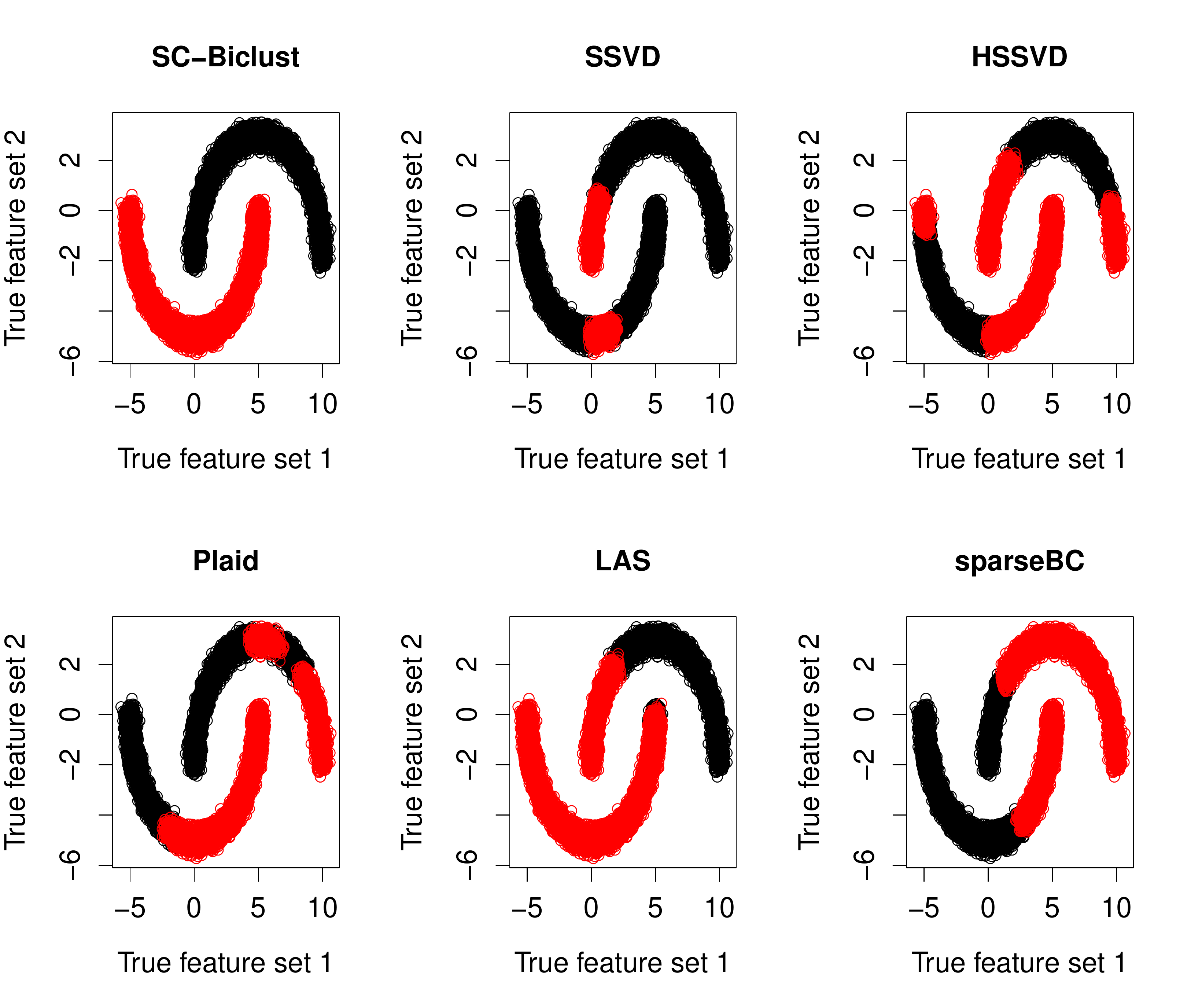}
\caption{\textit{Simulation example: non-spherical biclusters.} Each
  panel shows a plot of the second feature versus the first feature
  for a single simulation from the fourth simulation scenario. Note
  that the data forms two non-spherical clusters. Each panel shows the
  result of applying a biclustering method (specifically SC-Biclust,
  SSVD, HSSVD, Plaid, LAS, and sparse biclustering) to this data
  set. Observations that belong to the putative bicluster are labeled
  in red.} \label{F:sim_H}
\end{figure}

~\\
Most existing biclustering methods seek to maximize the Euclidean
distance between the center of the putative bicluster and the
remaining data values. This assumes that the biclusters are
approximately ``spherical'' (in the appropriate number of
dimensions). Although this assumption is reasonable in many
situations, it can cause these methods to fail if the assumption is
violated. See Figure \ref{F:sim_H} for an example of non-spherical
clusters in the case of two dimensions. Hierarchical clustering (with
single linkage) will do a better job of identifying clusters similar
to the clusters in Figure \ref{F:sim_H} than $k$-means clustering
(which also assumes that the clusters are spherical). One strength of
SC-Biclust is the fact that it can use clustering methods other than
2-means clustering to identify biclusters (see Section
\ref{S:var_biclust}). Thus, it is reasonable to expect that SC-Biclust
(with single linkage hierarchical clustering) will outperform
competing biclustering methods when the biclusters are non-spherical.

The purpose of this study was to provide an example where SC-Biclust
using hierarchical clustering can identify biclusters that existing
biclustering methods would fail to identify. Each 1200 $\times$ 75
data set was simulated as follows. For $1 \leq j \leq 25$:
\begin{align*}
  X_{i,2j} &= -2I(i \leq 500) + 5\sin(\theta_i + \pi I(i > 500)) +
  \epsilon_i \\
  X_{i,2j-1} &= 5I(i \leq 500) + 5\cos(\theta_i + \pi I(i > 500)) +
  \epsilon_i
\end{align*}
Here the $\epsilon_i$'s are iid $N(0,0.2)$ and the $\theta_i$'s are
iid $\text{Uniform}(0, \pi)$. For all $j>50$, the $X_{ij}$'s are
$N(0,1)$. Figure \ref{F:sim_H} shows the biclustering results from one
of the simulations. Each simulated data set was partitioned as
described in Section \ref{method_cv} to evaluate the reproducibility
of the biclusters.

\subsubsection{Variance Biclustering}
\begin{figure}
\centering
\includegraphics[scale=0.32]{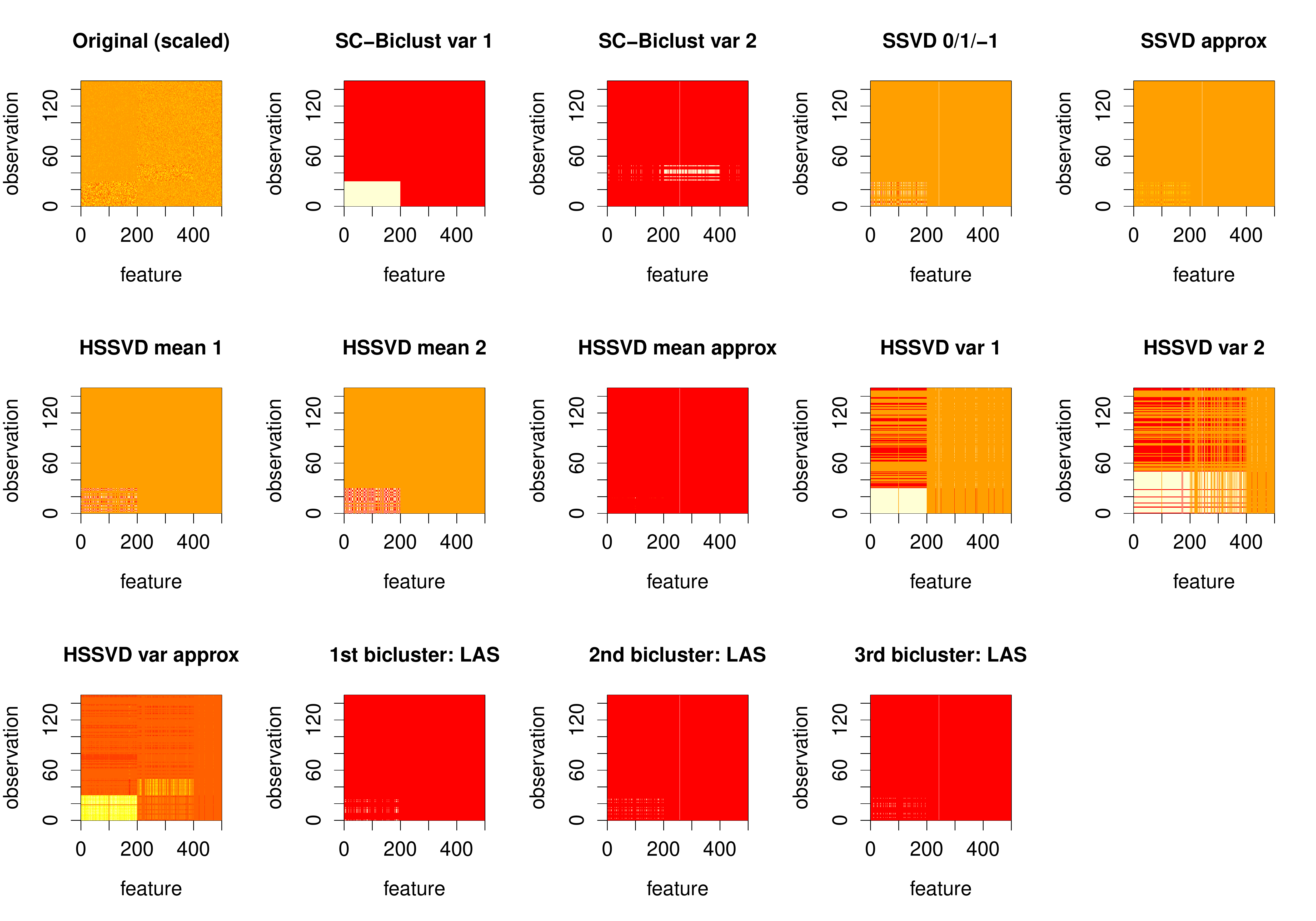}
\caption{\textit{Simulation example: variance biclustering.} This is
  an illustration of a single simulation from the fifth simulation
  scenario. The first panel shows a heat map of the
  (scaled) data. The two non-overlapping variance biclusters are on
  the bottom left corner. The remaining panels show the first two
  variance biclusters identified by SC-Biclust, result from SSVD and
  HSSVD, and the first three biclusters identified by LAS. The white
  regions correspond to the biclusters. For SSVD and HSSVD, both the
  0/1/-1 indicator matrix layers and the overall approximation
  matrices are plotted.} \label{F:sim_var}
\end{figure}

~\\
Another limitation of most existing biclustering methods is that they
are only capable of detecting biclusters whose mean values differ from
the data points not in the bicluster. In some situations, however, one
may wish to identify biclusters with higher (or lower) variance than
the data points not contained in the bicluster. As described in
Section \ref{S:var_biclust}, SC-Biclust can be modified to identify
biclusters with heterogeneous variance. The goal of this simulation is
to evaluate the ability of SC-Biclust to identify such biclusters. We
simulated data sets with two non-overlapping biclusters with
heterogeneous variances. Each simulated data set consisted of a 150
$\times$ 500 matrix with independent entries. The background entries
were all $N(1, 2)$. The first bicluster [1-30, 1-200] was generated as
$N(1, 15)$, and the second bicluster [31-50, 201-400] was generated as
$N(1, 5)$. Figure \ref{F:sim_var} shows the biclustering results from
one of the simulations. The prediction accuracy of the methods were
evaluated in the same way as the third simulation scenario, and no
reproducibility was assessed.

\subsubsection{Simulation Results}~\\
We simulated 100 data sets with the same structure for each simulation
scenario. Table \ref{T:sim_time} shows the average computing time for
each method for each simulation scenario. Tables \ref{T:sim_pred1} and
\ref{T:sim_pred2} show the prediction accuracy and the number of valid
biclusters, Table \ref{T:sim_cv} shows the reproducibility results for
simulations 1, 2, and 4, and Table \ref{T:sim_stop} shows the stopping
rule comparison for simulations 1 and 3.

SC-Biclust performed very well in the first simulation scenario. No
observations were misclassified across all 100 simulations and the
proportion of features that were misclassified was also very low. The
reproducibility of the biclusters identified by SC-Biclust was also
very good. The sparse biclustering method also produced good results,
except for the relatively high feature misclassification rate in the
reproducibility analysis. SSVD, HSSVD, and LAS tended to include
spurious features in the bicluster (as evidenced by their higher FPR),
and Plaid did not select any features for this simulation
scenario. The results of the second simulation scenario were
similar. Although the accuracy of SC-Biclust was lower when the
assumption of normality was violated, it produced a noticeably lower
error rate than competing methods (with the exception of LAS). More
importantly, SC-Biclust produced valid biclusters in all 100
simulations whereas SSVD, Plaid, and sparse biclustering frequently
failed to identify valid biclusters. Sparse biclustering tended to
produce good results when it identified biclusters in the data, but it
failed to detect any biclusters in 87 of the 100 simulations. LAS
identified valid biclusters in all 100 simulations with comparable
accuracy to SC-Biclust.

In the third simulation scenario, SC-Biclust identified both
biclusters with perfect accuracy in all the simulations. LAS also
identified the first bicluster with high accuracy but it tended to
include many spurious entries when identifying the second
bicluster. SSVD and HSSVD tended to identify bicluster 1+2 (combining
the two biclusters into one), and the performance of Plaid was
poor. The sparse biclustering method identified single biclusters, but
with very high false negative rates.

SC-Biclust had a much lower proportion of misclassified observations
in the fourth simulation scenario and excellent reproducibility. This
is not surprising, since the other biclustering methods assume that
the biclusters are spherical, and this assumption is violated for this
simulation. However, these results illustrate that SC-Biclust can be
used to identify biclusters in situations where existing methods will
fail.

In the fifth simulation scenario, SC-Biclust identified the first
variance bicluster with high accuracy. It usually detected the second
variance bicluster as well, although many of the entries were false
negatives. HSSVD tended to identify bicluster 1+2, with higher FNR and
FPR than SC-Biclust. The other methods performed poorly, which is not
surprising, since they are not designed to identify variance
biclusters.

In terms of computing time, SC-Biclust was generally significantly
faster than HSSVD and LAS and slightly faster than sparse biclustering
but slightly slower than SSVD and Plaid. However, in simulation 3,
SC-Biclust was significantly faster than all methods other than Plaid,
and SC-Biclust was faster than all methods other than SSVD on
simulation 5. On simulation 4, SSVD was noticeably slower than
SC-Biclust (as well as Plaid and sparse biclustering). Note that these
simulations use HSSVD without pre-specified rank, which increases
the computing time of this method.

In simulation 1, SC-Biclust correctly determined that 4 biclusters are
present in the data in 44\% of the simulations. It incorrectly
identified a 5th bicluster in 54\% of simulations and identified a
(non-existent) 6th bicluster in 2\% of simulations. HSSVD correctly
identified 4 biclusters in 54\% of simulations. The remaining methods
(namely LAS and sparse biclustering) consistently identified too many
biclusters. SSVD was not included in this comparison since it always
returns a single layer, and Plaid was not included since it did not
return any valid results for this simulation scenario. In simulation
3, SC-Biclust correctly determined that 2 biclusters are present in
the data in 99\% of the simulations. HSSVD correctly determined that 2
biclusters were present in 63\% of simulations, and Plaid determined
that 2 biclusters were present in 30\% of simulations. Again, LAS and
sparse biclustering always overestimated the number of biclusters, and
SSVD was not included in the comparison.

\begin{table}[ht]
\scriptsize
\centering
\caption{Comparison of computing times (average of 100
  simulations)}  \label{T:sim_time}
\begin{tabular} {l c c c c c}
\hline \hline
Algorithm & Simulation 1 & Simulation 2 & Simulation 3 & Simulation 4 & Simulation 5 \\
SC-Biclust & 0.416 sec & 0.42 sec & 0.79 sec &  4.93 sec & 1.91 min \\
SSVD & 0.28 sec & 0.62 sec & 0.39 sec & 37.34 sec & 51.41 sec\\
HSSVD & 1.25 min & 1.27 min & 1.28 min & 2.36 min & 5.05 min \\
Plaid & NA & 0.081 sec & 0.21 sec & 0.58 sec & NA \\
LAS & 12.50 sec & 1.27 min & 9.54 sec & 41.91 sec & 3.44 min \\
Sparse Biclustering & 0.85 sec & 0.99 sec & 23.95 sec &  4.62 sec & NA\\
\hline
\end{tabular}
\end{table}

\begin{table}[ht]
\scriptsize
\centering
\caption{Comparison of prediction accuracy: simulations 1, 2, and 4
  (average of 100 simulations)}  \label{T:sim_pred1}
\begin{tabular} {l c c c c}
\hline \hline
 & \multicolumn{4}{c}{Simulation 1:  primary bicluster identification}\\
Algorithm & Obs. misclassification rate & Feature FNR & Feature FPR &
Valid biclusters \\
SC-Biclust  & 0 & 0.15 & 0.0024  & 100 \\
SSVD  & 0.25 & 0 & 0.39 & 100 \\
HSSVD & 0.18 & 0 & 0.32 & 100 \\
Plaid & NA & NA & NA & 0 \\
LAS &  0.14 & 0.0021 & 0.38 & 100 \\
Sparse Biclustering & 0 & 0 & 0.0035 & 100 \\
\hline

& \multicolumn{4}{c}{Simulation 2:  departure from normality}\\
Algorithm & Obs. misclassification rate & Feature FNR & Feature FPR &
Valid biclusters \\
SC-Biclust  & 0.18 & 0.085 & 0.050 & 100 \\
SSVD  & 0.18 & 0.43 & 0.072 & 37 \\
HSSVD & 0.40 & 0.070 & 0.53 & 100 \\
Plaid & 0.28 & 0.33 & 0.13 & 62 \\
LAS &  0.20 & 0.017 & 0.27 & 100 \\
Sparse Biclustering & 0.00077 & 0.0058 & 0.0038 & 13 \\
\hline

& \multicolumn{4}{c}{Simulation 4:  non-spherical biclusters}\\
Algorithm & Obs. misclassification rate & Feature FNR & Feature FPR &
Valid biclusters \\
SC-Biclust  & 0.058 & 0 & 0 & 100\\
SSVD  & 0.41 & 0 & 0 & 100\\
HSSVD & 0.45 & 0 & 0 & 100\\
Plaid & 0.29 & 0.5 & 0 & 100\\
LAS &  0.12 & 0 & 0 & 100\\
Sparse Biclustering & 0.47 & 0.5 & 0 & 100\\
\hline

\end{tabular}
\end{table}

\begin{table}[ht]
\scriptsize
\centering
\caption{Comparison of prediction accuracy: simulations 3 and 5
  (average of 100 simulations)}  \label{T:sim_pred2}
\begin{tabular} {l c c c c}
\hline \hline
& \multicolumn{4}{c}{Simulation 3: sequential biclusters with overlap}\\
Algorithm & Identification & Entry FNR & Entry FPR & Valid biclusters\\
SC-Biclust layer 1 & Bicluster 1 100\% & 0 & 0 & 100 \\
SC-Biclust layer 2 & Bicluster 2 100\% & 0 & 0 & \\
SSVD  & Bicluster 1+2 100\% & 0.0048 & 0 & 100 \\
HSSVD mean layer 1 & Bicluster 1 26\%, Bicluster 1+2 74\% & 0.088 & 0.013 & 100\\
HSSVD mean layer 2 & Bicluster 1+2 100\% & 0.00017 & 0.00033 & \\
Plaid & Bicluster 1 98\% & 0.82 & 0.000073 & 98 \\
LAS layer 1 &  Bicluster 1 100\% & 0.022 & 0 & 100\\
LAS layer 2 &  Bicluster 2 100\% & 0.50 & 0.022 & \\
LAS layer 3 &  Bicluster 1 100\% & 1 & 0.064 & \\
Sparse Biclustering layer 1 &  Bicluster 1 85\%, Bicluster 2 15\% & 0.92 & 0.043 & 100\\
Sparse Biclustering layer 2 &  Bicluster 1 67\%, Bicluster 2 33\% & 0.87 & 0.026 & \\
Sparse Biclustering layer 3 & Bicluster 1 75\%, Bicluster 2 25\% & 0.91 & 0.071 & \\
\hline

& \multicolumn{4}{c}{Simulation 5: variance biclustering}\\
Algorithm & Identification & Entry FNR & Entry FPR & Valid biclusters\\
SC-Biclust layer 1 & Bicluster 1 99\%, Bicluster 2 1\% & 0.052 &  0.0000023 & 100 \\
SC-Biclust layer 2 & Bicluster 1 5\%, Bicluster 2 95\% & 0.76 & 0.0099 & \\
SSVD  & Bicluster 1 91\%, Bicluster 2 9\% & 0.78 & 0.0011 & 100\\
HSSVD mean layer 1 & Bicluster 1 100\% & 0.59 & 0.00013 & 100\\
HSSVD mean layer 2 & Bicluster 1 100\% & 0.20 & 0.00032 & \\
HSSVD variance layer 1 & Bicluster 1 100\% & 0.0016 & 0.16 & \\
HSSVD variance layer 2 & Bicluster 2 2\%, Bicluster 1+2 98\% & 0.23 & 0.35 & \\
Plaid & NA & NA & NA & 0\\
LAS layer 1 & Bicluster 2 100\% & 1 & 0.0036 & 100\\
LAS layer 2 & Bicluster 2 100\% & 1 & 0.0033 & \\
LAS layer 3 & Bicluster 2 100\% & 1 & 0.0024 & \\
Sparse Biclustering & NA & NA & NA & 0\\
\hline
\end{tabular}
\end{table}

\begin{table}[ht]
\footnotesize
\centering
\caption{Comparison of reproducibility (average of 100 simulations
  $\times$ 10 partitions)} \label{T:sim_cv}
\begin{tabular} {l c c c c }
\hline \hline
 & \multicolumn{4}{c}{Simulation 1:  primary bicluster identification}\\
Algorithm & Obs. misclassification rate & Feature FNR & Feature FPR & Feature misclassification rate\\
SC-Biclust & 0.11 & 0.18 & 0.041 & 0.14 \\
SSVD & 0.015 & 0.012 & 0.012 & 0.024 \\
HSSVD & 0.11 & 0.32 & 0.0075 & 0.13 \\
LAS & 0.061 & 0.15 & 0.023 &  0.19\\
Sparse Biclustering & 0.05 & 0.096 & 0.088 & 0.18 \\
\hline

 & \multicolumn{4}{c}{Simulation 2:  departure from normality}\\
Algorithm & Obs. misclassification rate & Feature FNR & Feature FPR & Feature misclassification rate\\
SC-Biclust & 0.29 & 0.12 & 0.093 & 0.19 \\
SSVD & 0.08 & 0.37 & 0.041 & 0.14 \\
HSSVD & 0.16 & 0.21 & 0.24 & 0.30 \\
Plaid & 0.37 & 0.29 & 0.15 & 0.19 \\
LAS & 0.048 & 0.21 & 0.010 &  0.19\\
Sparse Biclustering & 0.20 & 0.42 & 0.0030 & 0.093 \\
\hline

& \multicolumn{4}{c}{Simulation 4:  non-spherical biclusters}\\
Algorithm & Obs. misclassification rate & Feature FNR & Feature FPR & Feature misclassification rate\\
SC-Biclust  & 0.073 & 0 & 0 & 0 \\
SSVD  & 0.011 & 0 & 0 & 0 \\
HSSVD & 0.27 & 0.001 & 0 & 0.0005 \\
Plaid & 0.77 & 0.34 & 0.17 & 0.32 \\
LAS &  0.0047 & 0 & 0 & 0\\
Sparse Biclustering & 0.25 & 0 & 0 & 0 \\
\hline
\end{tabular}
\end{table}

\begin{table}[ht]
\scriptsize
\centering
\caption{Stopping rule comparison: simulations 1 and 3
  (average of 100 simulations)}  \label{T:sim_stop}
\begin{tabular} {l c }
\hline \hline
 & Simulation 1:  primary bicluster identification (4 biclusters are
 present in the data)\\
Algorithm & number of biclusters identified (\%) \\
SC-Biclust  & 4 (44\%) 5 (54\%) 6 (2\%) \\
HSSVD mean & 2 (5\%) 3 (39\%) 4 (54\%) 5 (2\%)\\
HSSVD var & 2 (100\%)\\
LAS &  8 (2\%) 9 (98\%) \\
Sparse Biclustering & 5 (1\%) 6 (99\%)\\
\hline

 & Simulation 3:  sequential biclusters with overlap (2 biclusters are
 present in the data)\\
Algorithm & number of biclusters identified (\%) \\
SC-Biclust  & 2 (99\%) 3 (1\%) \\
HSSVD mean & 2 (63\%) 3 (37\%)\\
HSSVD var & 2 (100\%)\\
Plaid & 1 (40\%) 2 (30\%) 3 (22\%) 4 (7\%) 5(1\%)\\
LAS &  7 (100\%) \\
Sparse Biclustering & 4 (100\%)\\
\hline
\hline

\end{tabular}
\end{table}

\subsection{Analysis of OPPERA data}
OPPERA is a prospective cohort study on Temporomandibular Disorders
(TMD), which are a set of painful conditions that affect the jaw
muscles, the jaw joint, or both. Both TMD-free participants and
chronic TMD patients were enrolled in the study. Each study
participant completed a quarterly questionnaire, and participants who
showed signs of first-onset TMD returned to the clinic for a formal
examination. The median follow up period was 2.8 years. The data set
contained 185 chronic TMD patients and 3258 initially TMD-free
individuals, 260 of whom developed TMD before the end of the
study. Among the TMD-free individuals, 521 did not complete any follow
up questionnaires and were excluded from the analysis. The remaining
2737 were used for survival analysis in the later sections, where
development of first-onset TMD is the event of interest. For a more
detailed description of the OPPERA study, see
\citeasnoun{slade2011study} or \citeasnoun{bair2013study}.

Three sets of possible risk factors for TMD were measured in OPPERA,
including autonomic measurements like blood pressure and heart rate
(44 total variables), psychosocial measurements like depression and
anxiety (39 total variables), and quantitative sensory testing (QST)
measurements (33 total variables) that evaluate participants'
sensitivity to experimental pain. See
\citeasnoun{fillingim2011potential}, \citeasnoun{greenspan2011pain},
and \citeasnoun{maixner2011potential} for more detailed descriptions
of these variables.

The SC-Biclust algorithm identified 3 significant biclusters within
the OPPERA data set. The first bicluster contained 30 measures of
autonomic function, the second bicluster contained 29 measures of
psychological distress, and the third bicluster contained 6 measures
of pain sensitivity. There were no overlap in the features selected in
the three biclusters. Thus, the biclusters identified by SC-Biclust
were consistent with the known structure of the data set. The
biclusters identified by the other methods did not correspond to the
three different types of measurements known to exist in this data
set. See Table \ref{T:oppera_biclust} for a summary of the results.

\begin{table}[ht]
\footnotesize
\caption{OPPERA: comparison of different biclustering algorithms} \label{T:oppera_biclust}
\centering
\begin{tabular} {l c c c}
\hline \hline
Algorithm (computing time) & Layer &  \multicolumn{2}{c} {Bicluster composition} \\
 & & $\#$ obs. (case; non-case) & $\#$ features (Auto; Psy; QST)\\
 \hline
SC-Biclust (2.59 min) & Layer 1 &  1561 (110; 1451) & 30 (30; 0; 0)\\
& Layer 2 & 998 (89; 909) & 29 (0; 29; 0) \\
& Layer 3 & 1619 (118; 1501) & 6 (0; 0; 6) \\

SSVD (21.28 sec) & Layer 1 & 3443 (185; 3258) & 98 (44; 22; 32) \\

HSSVD (12.47 min) & Mean 1 & 3443 (185; 3258) & 115 (44; 39; 32) \\
& Mean 2 & 3443 (185; 3258) & 116 (44; 39; 33) \\
& Var 1 & 3378 (184; 3194) & 109 (44; 36; 29) \\
& Var 2 & 3408 (185; 3223) & 111 (44; 36; 31) \\

Plaid  (14.08 sec) & Layer 1 & 68 (6; 62) & 23 (23; 0; 0) \\
& Layer 2 & 6 (1; 5)& 21 (21; 0; 0) \\
& Layer 3 & 23 (2; 21) & 21 (7; 14; 0) \\

LAS (14.66 min) & Layer 1 & 817 (33; 784) & 24 (24; 0; 0) \\
& Layer 2 & 638 (73; 565) & 43 (0; 23; 20) \\
& Layer 3 & 945 (78; 867) & 24 (24; 0; 0) \\
\hline
 \end{tabular}
 \end{table}

\begin{table}[ht]
\footnotesize
\caption{OPPERA: association between biclusters and chronic TMD} \label{T:oppera_chron}
\centering
\begin{tabular} {l c c c c c c}
\hline \hline
Algorithm  & \multicolumn{2}{c} {Bicluster 1} & \multicolumn{2}{c} {Bicluster 2} & \multicolumn{2}{c} {Bicluster 3} \\
& $\chi^2$ (df=1) & p-value & $\chi^2$ (df=1) & p-value & $\chi^2$ (df=1) & p-value \\
 \hline
SC-Biclust & 15.13 & $1.00 \times 10^{-4}$ & 33.75 & $6.26 \times
10^{-9}$ & 21.34 & $3.84 \times 10^{-6}$ \\
HSSVD var & 1.23 & 0.27 & 1.08 & 0.30 & NA & NA \\
Plaid  & 1.01 & 0.32 & 0.10 & 0.75 & 0.06 & 0.81 \\
LAS  & 3.41 & 0.065 & 55.27 & $1.05 \times 10^{-13}$ & 20.49 &
$6.01 \times 10^{-6}$ \\

\hline
 \end{tabular}
\end{table}

\begin{figure}
\centering
\includegraphics[scale=0.4]{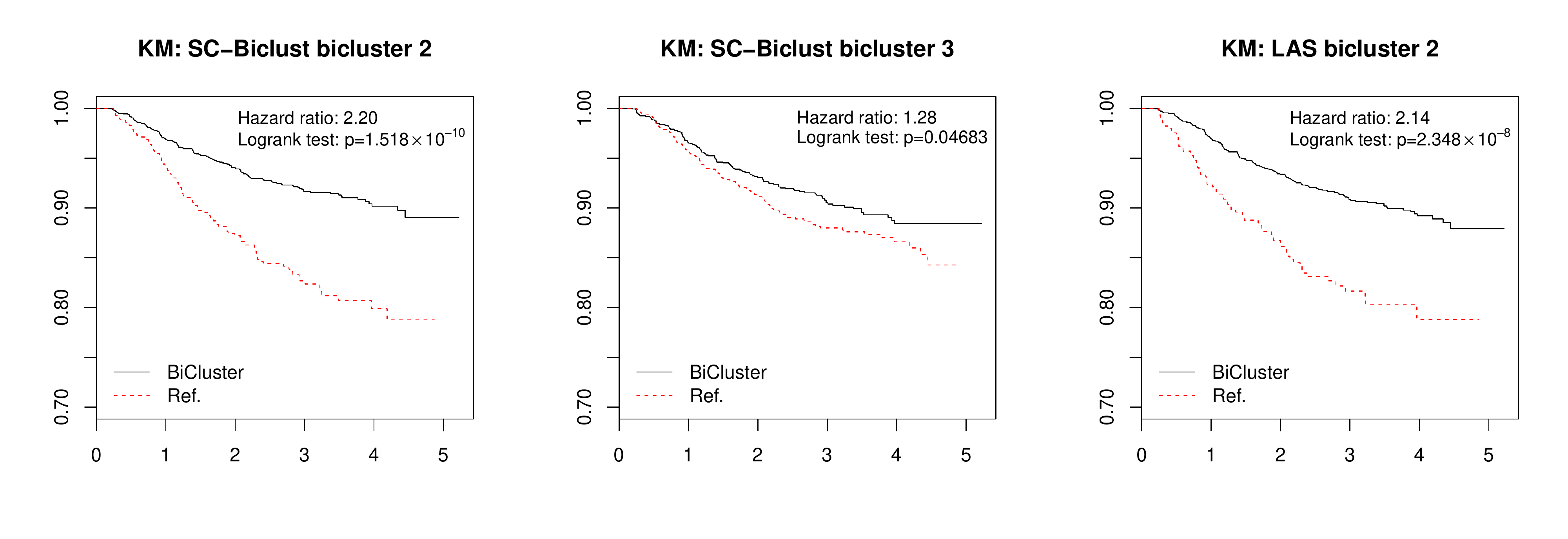}
\caption{\textit{OPPERA Kaplan-Meier plots.} The Kaplan-Meier plots
  showing the association between first-onset TMD and the biclusters
  identified by SC-Biclust (layer 2 and 3) and LAS (layer 2).} \label{F:oppera_surv}
\end{figure}

Membership in the biclusters identified by each method of interest was
evaluated as a possible risk factor for both chronic TMD and
first-onset TMD. (Subjects with chronic TMD were excluded from the
analysis for first-onset TMD.) The association between each bicluster
and chronic TMD is shown in Table \ref{T:oppera_chron}, and the
association between each bicluster and first-onset TMD is shown in
Table \ref{T:oppera_surv}. Kaplan-Meier plots for first-onset TMD for
selected biclusters are shown in Figure \ref{F:oppera_surv}. All three
biclusters identified by SC-Biclust were associated with chronic
TMD. The second bicluster was also associated with first-onset
TMD. The second and third biclusters identified by LAS were associated
with chronic TMD, and the second bicluster was also associated
with first-onset TMD. The remaining biclusters were associated with
neither chronic TMD nor first-onset TMD. SC-Biclust was faster than
HSSVD and LAS but slower than SSVD and Plaid. The sparse biclustering
algorithm failed to detect any biclusters.

\begin{table}[ht]
\footnotesize
\caption{OPPERA: association between biclusters and first-onset TMD
  (log-rank test)} \label{T:oppera_surv}
\centering
\begin{tabular} {l c c c c c c}
\hline \hline
Algorithm  & \multicolumn{2}{c} {Bicluster 1} & \multicolumn{2}{c} {Bicluster 2} & \multicolumn{2}{c} {Bicluster 3} \\
& Statistic (df) & p value & Statistic (df) & p value  & Statistic (df) & p value \\
 \hline
SC-Biclust & 2.72 (df=1) & 0.099 & 41.01 (df=1) & $1.52 \times 10^{-10}$
& 3.95 (df=1) & 0.047 \\
HSSVD var & 0.4 (df=1) & 0.53 & 0.26 (df=1) & 0.61 & NA & NA \\
Plaid  & 2.87 (df=1) & 0.090 & 0.42 (df=1) & 0.52 & 0.07 (df=1) & 0.80 \\
LAS  & 0.5 (df=1) & 0.48 & 31.18 (df=1) & $2.35 \times 10^{-8}$ & 1.71
(df=1) & 0.19 \\

\hline
 \end{tabular}
 \end{table}

\subsection{Analysis of a breast cancer gene expression data set}
The data set used in this section contains gene expression
measurements on 4751 genes from a total number of 78 breast cancer
subjects. The survival time of each subject is also available. See
\citeasnoun{van2002gene} for a more detailed description of this data
set.

The first bicluster identified by the SC-Biclust algorithm contains 16
subjects and 8 features. The first bicluster identified by the LAS
algorithm contains 16 observations and 1421 features. Interestingly,
the 16 observations identified by SC-Biclust and LAS are exactly the
same. The primary bicluster identified by the sparse biclustering
method contains 60 observations and 553 features. HSSVD method
identified 8 mean bicluster layers and 3 variance bicluster layers,
for which we will only study the primary mean layer. The Plaid method
failed to identify any biclusters within the data set, and the SSVD
method and the HSSVD variance identification did not produce valid
biclusters. Detailed biclustering results are provided in Table
\ref{T:friend1_surv}.

\begin{figure}
\centering
\includegraphics[scale=0.46]{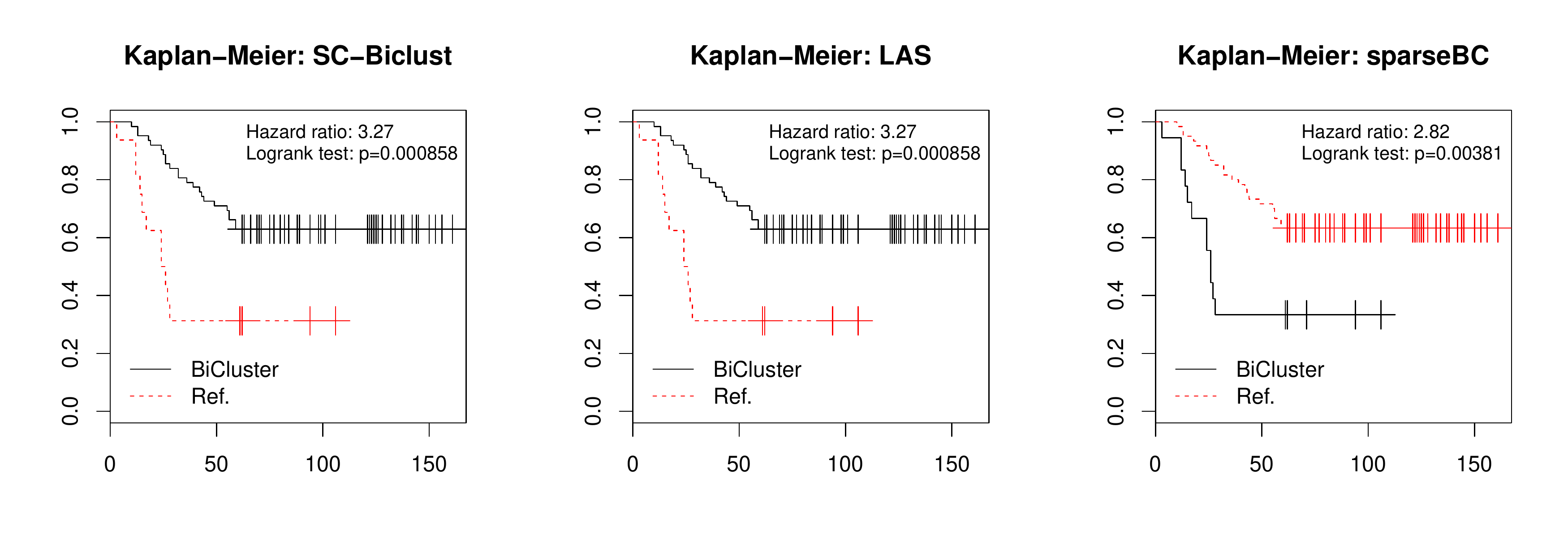}
\caption{\textit{Breast cancer gene expression Kaplan-Meier plot.} The
  Kaplan-Meier plots showing the association between survival and the
  biclusters identified by SC-Biclust, LAS, and sparse
  biclustering.} \label{F:friend1_surv}
\end{figure}

We tested the null hypothesis of no association between each putative
bicluster and survival using log rank tests. Table
\ref{T:friend1_surv} and Figure \ref{F:friend1_surv} show the
associations between survival and the biclusters identified by
SC-Biclust, HSSVD (mean layer only), LAS, and sparse biclustering. The
putative biclusters identified by SC-Biclust, LAS, and sparse
biclustering were associated with survival, but the putative bicluster
identified by HSSVD was not. The running time for SC-Biclust was also
significantly lower than the running time of the other methods.

\begin{table}
\centering
\footnotesize
\caption{Gene expression: Comparison of biclustering and survival
  analysis results.} \label{T:friend1_surv}
\begin{tabular} {l c c c c c}
\hline \hline
Algorithm  & Computing time & Obs. & Feature & \multicolumn{2}{c} {Score (log-rank) test} \\
 & & & & Statistic (df) & p value \\
 \hline
SC-Biclust  & 8.72 sec & 16 & 8 & 11.11 (df=1) & $8.58 \times
10^{-4}$ \\
HSSVD mean & 8.30 min$^*$ & 75 & 1046 &  0.42 (df=1) & 0.515 \\
LAS & 22.87 min & 16 & 1421 & 11.11 (df=1) & $8.58 \times 10^{-4}$ \\
Sparse Biclustering & 30.80 min & 60 & 553  & 10.2 (df=1) & 0.0014 \\
\hline
\end{tabular}
\end{table}

\subsection{Analysis of methylation data}
We applied SC-Biclust (and existing biclustering methods) to a
methylation data set comparing cancer patients with normal
patients. Methylation data was evaluated at 384 different
cancer-specific differentially methylated regions (cDMRs) for 138
normal samples and 152 cancer samples. Details of the data set are
described in \citeasnoun{hansen2011increased}, who reported that the
cancer samples had hypervariability in certain cDMRs compared to
controls.

We first applied the SC-Biclust algorithm to identify two mean
biclusters and then used the residual matrix for variance bicluster
identification, as described in Section \ref{S:var_biclust}. We chose
the top two variance biclusters for comparison with the other
methods. The HSSVD method identified two layers of mean biclusters and
six layers of variance biclusters. The Plaid method identified two
biclusters. The sparse biclustering method failed to detect any
biclusters. For the LAS method, we report the top three
biclusters. Comparison of the biclustering results are summarized in
Table \ref{T:methy}. The first mean bicluster identified by SC-Biclust
was strongly associated with cancer, as were both of the variance
biclusters. Indeed, we can see that the two variance biclusters
identified by the SC-Biclust algorithm contained cancer samples
exclusively. The other biclustering methods also identified biclusters
that were associated with cancer. It is interesting to compare the
variance biclusters identified by SC-Biclust to the variance
biclusters identified by HSSVD. SC-Biclust tends to identify small
variance biclusters that contain only cancer patients whereas HSSVD
tends to identify larger biclusters that are more heterogeneous. Note
that under the 0/1 transformation, SSVD and the mean layers of HSSVD
identified biclusters containing all of the observations. The running
time for SC-Biclust was greater than the running time of LAS and Plaid
but less than the running time of HSSVD.

\begin{table}[ht]
\footnotesize
\caption{Methylation: association between biclusters and
  cancer} \label{T:methy}
\centering
\begin{tabular} {l c c c c}
\hline \hline
Algorithm  & Layers  & Fisher's exact test  &  \multicolumn{2}{c} {Bicluster composition} \\
(Computing time) & & p-value & $\#$ obs. (cancer; normal) & $\#$ features \\
 \hline
SC-Biclust & Mean 1 &  $2.00 \times 10^{-11}$ & 115 (88; 27) & 274 \\
(5.74 min) & Mean 2 & 0.81 & 115 (59; 56) & 221 \\
& Var 1 & 0.00011  & 14 (14; 0) & 299 \\
& Var 2 & 0.00011 & 14 (14; 0) & 345 \\
 \hline
HSSVD & Mean 1 & NA & 290 (152; 138) & 243 \\
(9.02 min) & Mean 2 & NA & 290 (152; 138) & 261 \\
& Var 1 & $1.07 \times 10^{-14}$ & 190 (69; 121) & 235 \\
& Var 2 & 0.097  & 247 (124; 123) & 369 \\
& Var 3 & 0.13  & 249 (126; 123) & 373 \\
& Var 4 & 0.00021 & 262 (128; 134) & 376 \\
& Var 5 & 0.047  & 273 (139; 134) & 378 \\
& Var 6 & 0.047  & 273 (139; 134) & 379 \\
 \hline
Plaid & Layer 1 & $4.71 \times 10^{-5}$ & 13 (0; 13) & 104 \\
(1.22 sec) & Layer 2 &  0.031 &  6 (6; 0) & 80 \\
 \hline
LAS & Layer 1 & 0.45 & 53 (25; 28) & 232 \\
(3.98 min) & Layer 2 & $<2.2 \times 10^{-16}$ & 60 (60; 0) & 171 \\
& Layer 3 & $< 2.2 \times 10^{-16}$ & 58 (58; 0) & 100 \\
\hline
 \end{tabular}
 \end{table}

\section{Discussion}
Biclustering is an unsupervised learning algorithm that is a powerful
tool for studying HDLSS data. In this paper, we have proposed a general
framework for biclustering based on sparse clustering. We have
developed algorithms for heterogeneous mean and variance biclusters
as well as more complex structures that can be identified using
hierarchical clustering. The algorithms we described in this paper are
special cases of this framework, and similar methods can be developed
for other bicluster structures of interest.

The biclusters identified by SC-Biclust compared favorably with the
biclusters identified by competing methods for both the simulated and
real data sets. We believe that SC-Biclust has several other
advantages compared to existing biclustering methods. First, unlike
some other biclustering methods
\cite{lazzeroni2002plaid,tan2013sparse}, SC-Biclust does not assume
that all features in a bicluster have the same mean. This is a strong
assumption that is likely to be violated for many data sets. Indeed,
SC-Biclust does not even necessarily assume that the bicluster has
different means than the observations not in the bicluster. In
general, SC-Biclust can be applied given an arbitrary function whose
value increases as the ``difference'' between the bicluster and the
remaining observations increases and a method for maximizing this
function with respect to the observations. For example, as noted
earlier, SC-Biclust can be used to identify biclusters with
heterogeneous variance.

Note that singular value decomposition-based methods such as HSSVD can
also be used to denoise and approximate the data in addition to
bicluster detection. For this reason, although SC-Biclust
performs better in biclustering detection under the scenarios
considered in the paper, SVD-based methods can still be useful as a
preprocessing tool. The performance of SC-Biclust may be improved by
applying a preprocessing method first such as HSSVD, but this is beyond
the scope of this paper.

Second, SC-Biclust is noticeably faster than other biclustering
methods, particularly HSSVD and LAS. This was particularly true when
these methods were applied to high dimensional data. Thus, SC-Biclust
may be useful for Big Data problems where other methods are too
expensive computationally.

Finally, SC-Biclust provides a simple statistical test of the null
hypothesis that no biclusters exist. In general the problem of
determining if a bicluster (or any type of cluster) represents signal
or noise is difficult. Existing biclustering methods use various
stopping criteria to determine if a bicluster represents
signal. However, as demonstrated earlier, they frequently fail to
identify true biclusters or return putative ``biclusters'' that do not
actually exist. The stopping criteria used by SC-Biclust was generally
more accurate than these existing methods. Development of better
stopping criteria for biclustering methods is an important area for
future research.

One limitation of using the beta distribution to approximate the
null distribution of the weights for SC-Biclust is the fact that it
requires the assumption that the BCSS of the features are independent
of one another. This is unlikely to be satisfied if the features
themselves are correlated, which is likely to be true in most real
world data sets of interest. Fortunately our experience suggests that
the correlations of the BCSS's tend to be modest even when the
original features are strongly correlated with one another, so
SC-Biclust tends to be robust against violations of this
assumption. In particular, when multiple biclusters exist in a data
set, this assumption is necessarily violated, which was the case for
most of our simulation examples. Nevertheless SC-Biclust generally did
not identify spurious biclusters in these simulations. However, if
there is evidence of strong correlations among the BCSS's, it may be
preferable to approximate the null distribution of the weights using
the nonparametric permutation method.

It is interesting to compare the results of SC-Biclust and HSSVD for
variance biclustering. In the examples considered in this manuscript,
SC-Biclust tended to identify smaller, more homogeneous biclusters
whereas HSSVD tended to identify biclusters that were larger and more
heterogeneous. It is unclear if this result is true in general or if
it is merely an artifact of these particular data sets. In any event,
one potential advantage of HSSVD for this problem is that SC-Biclust
mail be less likely to detect ``small'' biclusters than HSSVD. It is
possible that the method used by SC-Biclust to identify variance
biclusters could be improved. The identification of variance
biclusters is a relatively new topic and an important area for future
research.

\appendix

\makeatletter   
 \renewcommand{\@seccntformat}[1]{APPENDIX~{\csname the#1\endcsname}.\hspace*{1em}}
 \makeatother

\bibliographystyle{ECA_jasa}
\bibliography{reference}

\end{document}